\documentclass{article}
\usepackage{spconf,amsmath,graphicx}
\title{Referenceless Performance evaluation of audio source separation using deep neural networks}
\name{Emad M. Grais$^{1,3}$, Hagen Wierstorf$\,^{1}$\sthanks{The first two authors started this work when they were researchers at Centre for Vision, Speech and Signal Processing, University of Surrey, UK. They contributed equally to this work. Hagen Wierstorf is now with audEERING GmbH, Gilching, Germany.}, Dominic~Ward$^{1}$, Russell~Mason$^{2}$, Mark~D.~Plumbley$^{1}$}
\address{$^1$Centre for Vision, Speech and Signal Processing, University of Surrey, UK \\
$^2$Institute of Sound Recording, University of Surrey, UK \\
$^3$Electronics, Communications, and Computer Engineering Department, Helwan University, Egypt}
\begin{document}
%\ninept
%
\maketitle
\begin{abstract}
Current performance evaluation for audio source separation depends on comparing the processed or separated signals with reference signals. Therefore, common performance evaluation toolkits are not applicable to real-world situations where the ground truth audio is unavailable. In this paper, we propose a performance evaluation technique that does not require reference signals in order to assess separation quality. The proposed technique uses a deep neural network (DNN) to map the processed audio into its quality score. Our experiment results show that the DNN is capable of predicting the sources-to-artifacts ratio from the blind source separation evaluation toolkit \cite{vincent:06:pmi} without the need for reference signals.
%Current performance evaluation for audio source separation techniques depend on comparing the processed or separated signals with reference signals. In these techniques, the reference signals are assumed available which restricts the practicality of such evaluation techniques in real-world scenarios since the reference signals usually are not available. In this paper, we propose a performance evaluation technique that does not require the reference signals. The proposed technique uses a deep neural network (DNN) to evaluate the quality of the processed audio signal by mapping the input processed signal to a quality score. Our experiment results show that DNN is capable of predicting the source-to-artifacts ratio from the blind source separation evaluation toolkit without the need for the reference signals. 
\end{abstract}
\section{Introduction}
\label{sec:intro}
Audio source separation aims to separate one or more target audio sources from mixture signals \cite{Virtanen:07:msssbnmfwtcasc,Stefan:17:imssbdnntdanb}. The separated sources often contain distortions, artifacts, and unwanted signals from the other sources in the mixtures. An evaluation of the quality of the separated sources is essential to guide development of separation algorithms or to select the most suitable algorithm for a given mixture signal or application type. This requires either perceptual evaluation where experienced listeners judge the quality of the estimated sources according to different perceptual attributes \cite{emiya:11:soqaass,ward:18:bepppsvs,hagen:17:pessrm,Cartwright:16:fecpae,Coleman:18:pebssobap,Cano:16:eqsssahpqm,itu:15:msaiqlas}, or objective metrics that can estimate the proportion of distortions, artifacts, or interference present in the separated sources, by comparing these with the reference clean sources \cite{vincent:06:pmi,huber:06:pemoq}. 
%Many quality assessment techniques have been proposed to evaluate the quality of the separated audio signals \cite{vincent:06:pmi,emiya:11:soqaass,Coleman:18:pebssobap,Cano:16:eqsssahpqm,huber:06:pemoq,Cartwright:16:fecpae,itu:15:msaiqlas}. Most of these techniques depend on comparing the output signals of a source separation system with their corresponding references or clean signals.  

In experimental situations, the reference sources are usually available for use in evaluating the performance of a certain source separation approach. However, for practical applications of source separation, the mixtures are available but the separate original sources (the reference signals) are not. Without these reference sources being available, the most common objective metrics cannot be employed, and the only way to evaluate the quality of the separated sources is to ask listeners to give scores for the quality of the separated sources. Using listeners to evaluate the quality of the separated sources is time consuming, and often unfeasible, and hence an automated system of evaluating the quality of the separated signal using neither listeners nor reference signals would be preferable. Such an automated referenceless evaluation method could be useful, for example, for selecting the most appropriate source separation algorithms for soloing or karaoke applications for each song, or automatically evaluating whether the separated signals are of sufficient quality or whether extra work is needed to further improve the quality of the separated signals using post-processing or additional separation techniques, e.g. \cite{emad:13:stpemenbscss,Williamson:14:tsaipqss}.

The concept of referenceless quality evaluation for processed signals has been introduced in many signal processing domains, including the perceptual evaluation of image enhancement approaches \cite{hossein:17:lpie}, and evaluating the quality and intelligibility of speech signals \cite{szuwei:18:qnetenisqambblstm,Spille:18:psidnn}. In this paper we propose a referenceless evaluation method to evaluate the quality of the separated audio sources without using the reference sources. The main idea of the proposed method is to train a deep neural network (DNN) to map the estimated separated sources to the output of a reference-based evaluation metric. The metric used in this paper is the Sources-to-Artifacts Ratio (SAR) from the Blind Source Separation Evaluation (BSS Eval) toolkit \cite{vincent:06:pmi}. SAR is selected as a case study, but it is intended that the proposed method will be used for other objective metrics, or the results of subjective judgments.

The DNN is first trained to map the separated signals from one or more source separation algorithms to their SAR scores. SAR in the training stage of the DNN is calculated  by using the reference signals of each source. The trained DNN is then used to estimate the SAR for separated sources without using any reference signals. %From the estimated SAR, we can know whether the separated sources are good enough or more processes are needed to further improve the quality of the separated sources.  

We consider three different scenarios of using DNNs to estimate the SAR values. The first scenario is to evaluate how well a DNN can predict the SAR results for the same single source separation algorithm for which it is trained: we refer to this scenario as a \textit{within-algorithm test}. The second scenario is to evaluate how well a DNN can predict the SAR results for a range of separation algorithms when trained using data from that same set of separation algorithms: we refer to this scenario as an \textit{across-known-algorithms test}. The third scenario is to evaluate how well a DNN can predict the SAR results for a range of separation algorithms when trained using data from a different set of separation algorithms: we refer to this scenario as an \textit{across-unknown-algorithm test}. 
%
%train and test the DNN on separated sources from a certain source separation method. The second scenario is to train the DNN on the separated sources using one source separation technique and test it on separated sources using other source separation techniques. The third scenario is to train the DNN on separated sources from many source separation techniques and test it on separated sources from different set of source separation techniques. Our experimental results show that DNN is capable of finding good estimates for SAR in the three scenarios.

\section{The Blind Source Separation Evaluation toolkit}
The Blind Source Separation Evaluation (BSS-Eval) toolkit \cite{vincent:06:pmi} is the most frequently used tool for evaluating source separation algorithms. BSS-Eval decomposes the error between the reference/target source and the extracted/separated source into a target distortion component reflecting spatial or filtering errors, an artifacts component pertaining to artificial noise, and an interference component associated with the unwanted sources. The salience of these components is quantified using three energy ratios: source Image-to-Spatial distortion Ratio (ISR), Sources-to-Artifacts Ratio (SAR), and Source-to-Interference Ratio (SIR). A fourth metric, the Source-to-Distortion Ratio (SDR), measures the global performance (all impairments combined). Computing these metrics depends mainly on comparing the reference signals and their corresponding estimated signals from the source separation system for each source. Without the reference sources, the BSS-Eval toolkit cannot provide information regarding the quality of the estimated sources.   

\section{Deep neural network for referenceless SAR prediction}
\begin{figure}[t]
 \centering
% \scalebox{0.82}
%{
    \includegraphics[width=0.75\linewidth,height=6.5cm]{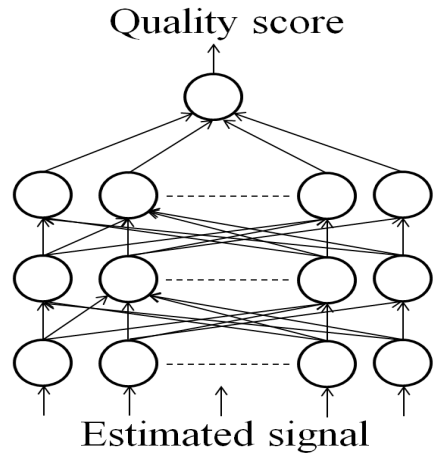}
    \caption{\footnotesize{The deep neural networks structure that we use in this work. The input is the estimated separated signal and the output is its corresponding quality score.}}
 \label{fig:dnn}
\end{figure}

In this paper we use a deep neural network to predict the BSS-Eval SAR scores from the output signals of a source separation system. The DNNs we use are fully connected feed forward neural networks as shown in Fig. \ref{fig:dnn}. SAR was selected as a case study: it has been shown to be an indicator of the magnitude of perceptual artifacts in the separated signals \cite{ward:18:bepppsvs,hagen:17:pessrm}.
% It has been shown that SAR is an important measurement the for remixing applications of the separated sources \cite{hagen:17:pessrm}. 

The DNN is trained to map the extracted features of the separated sources to their corresponding SAR values. In this training stage of the DNN, we assume the reference signals are available. Given the reference or clean signals and their corresponding estimated signals from the source separation technique, the SAR is calculated using BSS-Eval \cite{vincent:06:pmi}. We extract features from the separated sources and use these features as input to the DNN. The features we use in this work are the mel-frequency spectrogram (MFS), which are calculated by converting the spectrograms of the estimated signals to a mel-frequency scale with 128 frequency channels. The training of the DNN parameters is done by minimizing the mean-square-errors between the estimated SAR values from the DNN and their corresponding calculated SAR values using BSS-Eval.  
% Predict a single SAR value from 40*128=5120 mel spectrogram values 
% We consider the SAR as the quality score of the separated sources.
%[Hagen: should be spectrogram as we don't apply any discrete cosine transform on it. I guess our main motivation to switch from the spectrogram to a mel-spectrogram was the reduction of frequency channels]

The trained DNN is then used to estimate the SAR values for a new set of separated sources without using the reference signals. The MFS features are extracted from the separated sources and fed to the trained DNN to estimate the SAR values of the input features.

%email SAR DNN predictions, first results
%Here are some stats on what we did:
%* Predict a single SAR value from 40*128=5120 mel spectrogram values
%* Feed-Forward model with 3 hidden layers
%* Middle layer size: 500
%* SAR clipped clipped to [-20, 30]
%* No data normalization for the model
%* Trained on 70% of all SiSEC songs, tested on the remaining 30%
%* Used different algorithm for test and train (problem: NUG1, NUG2, ...)
%* Here the last lines of the training output:

\section{Experiments}
\label{sec:method}
We undertook a pilot study to predict the sources-to-artifact ratio (SAR) as provided by BSS-Eval. The audio data and the source separation algorithms were taken from the SiSEC-2016-MUS-task challenge \cite{Liutkus:17:ssec}. The data consists of 100 stereo songs, though four of them are corrupted so were removed. Each song is a mixture of vocals, bass, drums, and other musical instruments. The SiSEC-2016-MUS-task involved separating these four sources from each song in the dataset. In total, 24 different source separation algorithms with differing performance were submitted to this challenge. The following submitted source separation algorithms are blind source separation algorithms: DUR \cite{Durrieu:12:ammrpemass}, KAM \cite{Liutkus:15:saslkam}, OZE \cite{Ozerov:12:gffhpias}, RAF \cite{Rafii:13:rpet}, JEO \cite{Jeong:17:svsrpca}, and HUA \cite{Huang:12:svsmrrp}, and the following submitted algorithms are supervised source separation algorithms using deep neural networks: STO \cite{Stoter:16:cfmuss}, UHL \cite{Stefan:17:imssbdnntdanb}, NUG \cite{arie:16:masswdnn}, CHA \cite{chandna:17:massudcnn}, GRA \cite{Emad:16:scassdnne}, and KON \cite{site:sisec17}. The separated signals using the Ideal Binary Mask (IBM) \cite{Liutkus:17:ssec} are also included in this data. More details about each algorithm can be found in the SiSEC-2016 website \cite{site:sisec17}. These source separation algorithms produced separated signals with a wide range of SAR values (from $-10$\,dB to $20$\,dB). 

In our experiments we aimed to predict the SAR for the vocal separated from each song for all the source separation algorithms that were submitted to this challenge. We tested three different scenarios of varying difficulty:
\begin{itemize}
  \item Test 1: The DNN model was used to predict the SAR for the source separation algorithm for which it had been trained. We call this test a \textit{within-algorithm test}. This was conducted separately for each separation algorithm to examine any algorithm-dependence in the results.
  \item Test 2: The DNN model was trained using data from all 24 source separation algorithms simultaneously, then used to predict SAR values of each of the 24 source separation algorithms. We call this test an \textit{across-known-algorithms test}. 
  \item Test 3: The DNN model was trained using data from 17 source separation algorithms simultaneously, then used to predict SAR values for 7 source separation algorithms not used in the training. We call this test an \textit{across-unknown-algorithm test}.
\end{itemize}
  
The 96 available songs (non-corrupted) from SiSEC-2016 dataset were split into 67 training songs and 29 test songs, all processed by the algorithms used in the tests. As the perceptual quality varies over time for musical signals, the SAR was calculated every 117 milliseconds (ms) over a time window of 464 ms on an 116 seconds (s) excerpt of every song. 
%starting at 10 s for Test 1 and 2, and starting at 0 s for Test 3 [Hagen: Mark asked for the reason that we used 0s for test 3, but 10s for test 1 and test 2
%We actually started with test 3 and 0s, but realized that some songs included no vocals at its beginning and it would be saver to start later in the song. Then we changed it for test 1 and test 2.]. 
The goal of the trained DNNs was to predict the time-varying SAR for every song and source separation algorithm in the test data set. The DNNs were deep fully connected feed forward networks as shown in Fig. \ref{fig:dnn}, consisting of three hidden layers using a rectified linear unit (ReLU) activation function for all but the last layer, which used a linear activation function. The number of nodes in each hidden layer was 500. The input features were calculated as follows: the stereo inputs were converted to mono by taking the average between the two channels; the spectrogram was calculated and converted to mel-frequency spectrograms (MFS) with 128 frequency channels. We stacked 40 neighbour MFS frames to form the inputs of the DNN with dimension $40\times128=5120$ MFS values, where 40 is the number of stacked frames, and each frame contains 128 frequency bands. 

To evaluate how well the DNNs could predict the SAR values without using the reference signals, we compared the estimated SAR as output from the DNNs with the SAR values calculated from the BSS-Eval toolkit using the reference signals; the average absolute error and the correlation between these were used to evaluate the performance of the DNN accuracy. 

%WHAT IS FIG. \ref{fig:exp2} FOR? WHICH SCENARIO?
%HOW DID WE HANDEL THE STEREO INFORMATION IN THE INPUT FEATURES? Hagen: we ignored stereo by converting the wav files to mono with untwist during loading. Untwist converts to mono by taking the average.
%%%%%%%%%%%%%%%%%%%%%%%%%%%%%%%%%%%%%%%%%%%%%%%%%%%%%%%%%%%%%%%%%%%%%%%%%%%%%%%%
\section{Results}
\label{sec:results}
%--%--%--%--%--%--%--%--%--%--%--%--%--%--%--%--%--%--%--%--%--%--%--%--%--%--%-

Table \ref{all_sisec2016} shows the mean absolute error and the mean correlation between the referenceless estimated SAR values using DNNs and the calculated SAR using BSS-Eval with reference signals (reference SAR) for the three scenarios (Test 1 to Test 3). 

\begin{table}[t]
\centering % used for centering table
\scalebox{0.8}
{
\begin{tabular}{||c | c c | c c | c c||} % centered columns (4 columns)
\hline\hline %inserts double horizontal lines
        & \multicolumn{2}{|c|}{Test1} & \multicolumn{2}{|c|}{Test2} & \multicolumn{2}{|c||}{Test3}  \\ 
[0.5ex] 
%heading
Algorithm  & Error & Corr. & Error & Corr. & Error & Corr.  \\
% inserts table         
\hline % inserts single horizontal line
CHA  & 1.2 & 0.82 & 1.5 & 0.83 & 0.7 & 0.89 \\
GRA2 & 1.4 & 0.87 & 1.5 & 0.86 & 1.3 & 0.92 \\
GRA3 & 1.3 & 0.80 & 1.6 & 0.81 & 1.7 & 0.89 \\
IBM  & 1.3 & 0.90 & 2.9 & 0.86 & 3.1 & 0.93 \\
JEO1 & 0.8 & 0.89 & 1.3 & 0.76 & 0.9 & 0.89 \\
KAM1 & 1.2 & 0.83 & 1.2 & 0.79 & 0.9 & 0.87 \\
KAM2 & 0.9 & 0.81 & 1.0 & 0.75 & 0.6 & 0.85 \\
KON  & 1.3 & 0.90 & 1.3 & 0.88 & 1.3 & 0.92 \\
NUG1 & 1.4 & 0.89 & 1.1 & 0.88 & 0.5 & 0.95 \\
NUG2 & 1.3 & 0.89 & 1.1 & 0.88 & 0.5 & 0.96 \\
NUG3 & 1.4 & 0.89 & 1.2 & 0.89 & 0.8 & 0.95 \\
OZE  & 1.0 & 0.72 & 1.1 & 0.73 & 0.9 & 0.80 \\
RAF1 & 0.9 & 0.75 & 1.3 & 0.72 & 1.2 & 0.78 \\
STO1 & 1.1 & 0.90 & 1.0 & 0.87 & 0.5 & 0.94 \\
UHL3 & 1.5 & 0.86 & 1.8 & 0.85 & 1.5 & 0.93 \\
%\hline
NUG4 & 1.5 & 0.89 & 1.2 & 0.89 & 1.6 & 0.92 \\
UHL2 & 1.5 & 0.84 & 1.7 & 0.85 & 1.5 & 0.90 \\
\hline
DUR  & 1.2 & 0.75 & 1.7 & 0.72 & 3.7 & 0.74 \\
HUA  & 0.8 & 0.66 & 1.1 & 0.61 & 4.4 & 0.30 \\
JEO2 & 0.8 & 0.95 & 1.1 & 0.93 & 1.6 & 0.93 \\
RAF2 & 1.0 & 0.77 & 1.1 & 0.73 & 1.4 & 0.70 \\
RAF3 & 1.0 & 0.82 & 1.4 & 0.78 & 2.0 & 0.79 \\
STO2 & 1.1 & 0.90 & 1.0 & 0.88 & 1.1 & 0.88 \\
UHL1 & 1.4 & 0.85 & 1.3 & 0.86 & 1.5 & 0.86 \\
\hline %\\ %[1ex]%inserts single line
\end{tabular}
}
\\ [1ex]
\caption{\footnotesize{The mean absolute error in dB and the mean correlation between the referenceless estimated SAR values using DNNs and the calculated SAR using BSS-Eval with reference signals (reference SAR) for each source separation algorithm. The horizontal line separates the algorithms used for training (above the line) and those used for testing (below the line) in Test 3.}} % title of Table
\label{all_sisec2016} % is used to refer this table in the text
\end{table}
\subsection{Test 1: the within-algorithm test}

Test 1 was intended to be a case where a DNN could be trained individually for a given separation algorithm, and hence should give the most favourable results as the DNN is customised for a single case. For this, we independently trained 24 DNNs: one for each source separation algorithm. Each DNN in this case was used to estimate the SAR for the separation algorithm for which is was trained. The same set of training songs and the same set of test songs was used for each algorithm, with no overlap between the two sets of songs. The error in the predictions was calculated as the difference between the predicted SAR from each DNN, and the reference SAR for the same separated signal. The mean absolute error between the predicted and reference SAR was $1.2$\,dB, and ranged from $0.8$\,dB to $1.5$\,dB for each separation algorithm. The correlation between the predicted and measured SAR ranged from $0.66$ to $0.95$ for each algorithm, with an average over the 24 algorithms of $0.84$.

Compared to the range of SAR values of $-10$\,dB to $20$\,dB, the mean absolute error of $1.2$\,dB represents 4\% of the range. This suggests that the SAR values estimated without using a reference could be used to discriminate between the performance of some combinations of algorithm and song. However, it may not be able to discriminate between the average results of some of the algorithms in the SiSEC-2016-MUS-task \cite{Liutkus:17:ssec}, and hence further refinement is required. 
%
% The mean absolute error across all algorithms and test songs was $2.4$\,dB, and 
% ranging from $0.66$ for HUA to $0.95$ for JEO2. %The mean absolute error was calculated as the mean value of the calculated SAR values from BSS-Eval within the song minus the mean values of the estimated SAR using DNNs within the song. The mean absolute prediction error of per song SAR was calculated as the mean values of the deference between the calculated SAR from BSS-Eval and the estimated SAR using DNNs within the song. In other words, The mean absolute error is the difference of the means, while absolute prediction error is the mean of the differences. - RM: I don't follow what you mean here. Leave it out?
%
%[THE ERROR DOES NOT GIVE GOOD INFORMATION, I THINK PERCENTAGE ERROR CAN BE BETTER BUT ALSO DISAPPOINTED. THE RANGE OF SAR IS USUALLY less than 20 dB AND 2.4 dB ERROR MEANS WE HAVE AT LEAST 10\% ERROR, i THINK CORRELATION IS MORE IMPORTANT SINCE THE QUALITY SCORE CAN HAVE DIFFERENT RANGE OF VALUES FROM ONE TOOLKIT TO ANOTHER.]
%
%--%--%--%--%--%--%--%--%--%--%--%--%--%--%--%--%--%--%--%--%--%--%--%--%--%--%-
\subsection{Test 2: the across-known-algorithms test}
Test 2 was intended to be a case where a single DNN was trained using a set of separation algorithms, and this used to attempt to predict the results of any separation algorithm included in its training set. This requires a more generalised set of predictions compared to Test 1, and hence was intended to be a more challenging test. The single DNN was trained using the same training set of songs employed in Test 1, though this time using the results from all 24 source separation algorithms. The trained DNN was then used to evaluate the separated vocal signals from the test set songs individually for each of the same 24 source separation algorithms. The results are shown in Table \ref{all_sisec2016}: the mean absolute error between the predicted and reference SAR was $1.4$\,dB, and ranged from $1.0$\,dB to $2.9$\,dB for each separation algorithm. The correlation between the predicted and measured SAR ranged from $0.61$ to $0.93$ for each algorithm, with an average over the 24 algorithms of $0.82$.
% The mean absolute error across all algorithms and test songs was $2.7$\,dB, and the mean absolute prediction error per song was $1.4$\,dB.
\begin{figure}[t]
    \includegraphics[width=1\columnwidth]{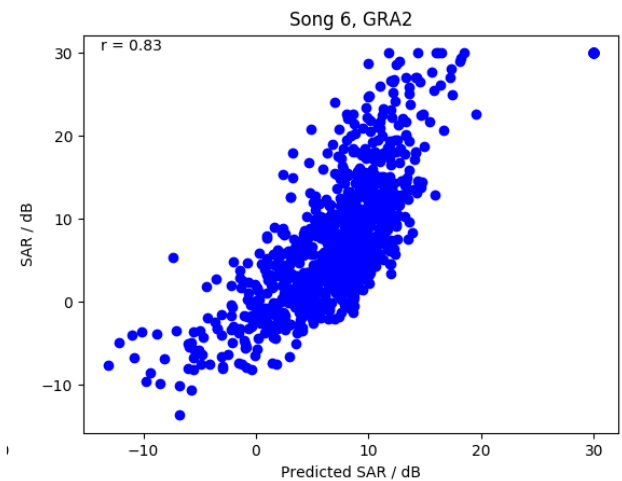}
    \caption{\footnotesize{The correlation between the estimated and reference SAR values for a song separated by source separation algorithm GRA2.}}
    \label{fig:exp2}
\end{figure}

As an example of the correlation between the estimated and actual SAR results, Fig. \ref{fig:exp2} shows the correlation between the estimated and reference SAR values for a song separated by source separation algorithm GRA2. As can be seen from the figure, the estimated SAR values are highly correlated with the reference SAR.

Compared to the range of SAR values of $-10$\,dB to $20$\,dB, the mean absolute error of $1.4$\,dB represents nearly 5\% of the range. Though the performance is less accurate for this more challenging test, even the worst-case mean absolute error of $2.9$\,dB indicates that the referenceless SAR prediction could be used to discriminate between the performance of some combinations of algorithm and song, but again further refinement is required. 
%Fig. \ref{fig:exp2} shows an example of the estimated SAR values using DNN for a song separated by source separation algorithm GRA2 and the corresponding SAR calculated by BSS-Eval (reference SAR). The figure also shows the correlation between the estimated and reference SAR values. As can be seen from the figure, the estimated SAR values are highly correlated with the reference SAR. 
%I THINK THIS FIGURE SHOULD BE REMOVED. THE DEFERENCE BETWEEN THE REFERENCE SAR AND THE ESTIMATED IS $>$ 10dB IN LARGE PERIOD OF TIME. RM: it looks like this is usually when the actual SAR is >20dB. In practice this might not be a major problem because the SAR is good enough from either measure?
%--%--%--%--%--%--%--%--%--%--%--%--%--%--%--%--%--%--%--%--%--%--%--%--%--%--%-
\subsection{Test 3: the across-unknown-algorithm test}

Test 3 was intended to be a case where a single DNN was trained using a set of separation algorithms, and this used to attempt to predict the results of any separation algorithm, including those not included in its training set. This requires further generalisation of the results, to both songs and algorithms outside of the training set, and is the most challenging of the tests used. For this, the first 17 source separation algorithms in Table \ref{all_sisec2016} were used for training and validation, and the last 7 algorithms (separated by a horizontal line in Table \ref{all_sisec2016}) were used for testing; the training and testing were again undertaken using a separate sets of songs. In addition, the DNN was tested separately for each source separation algorithm using solely the songs from the test set, with the results shown in Table \ref{all_sisec2016}). The mean absolute error between the predicted and reference SAR was $2.3$\,dB, and ranged from $1.1$\,dB to $4.4$\,dB for each separation algorithm in the test set, and from $0.5$\,dB to $3.1$\,dB for each separation algorithm in the training set. The average correlation between the predicted and measured SAR time series was $0.74$, with a range of $0.3$ to $0.93$ for the test set and $0.78$ to $0.96$ for the training set.

As expected, the performance was less accurate for this test, though the worst-case error would still allow discrimination between some combinations of algorithm and song. 
%The mean absolute error across all algorithms and test songs was $3.1$\,dB, and the mean absolute prediction error per song was $2.3$\,dB. The average correlation between the predicted and measured SAR time series was $0.74$.
%%%%%%%%%%%%%%%%%%%%%%%%%%%%%%%%%%%%%%%%%%%%%%%%%%%%%%%%%%%%%%%%%%%%%%%%%%%%%%%%

%
%\section{Discussion}
%\label{sec:discussion}
%
%Things to look at:
%\begin{itemize}
%    \item Why did HUA perform particular worse in all experiments?
%   \item Introduce proper data normalisation of input and output data of the
%        DNN model
%    \item Predict other metrics like SIR and SDR
%    \item Try to predict the error signal used by BSS eval to calculate those
%        metrics instead. This would have several advantages, e.g. the time
%        resolution is not fixed.
%    \item Extend to SiSEC 2018 data set
%    \item Test the influence of the algorithm of the source separation algorithm on
%        the prediction for unknown algorithms. For example, is there an advantage
%        if a unknown algorithm has a similar structure then a known one, or will it
%        also work for completely different algorithms (e.g. non-DNNs).
%\end{itemize}

%%%%%%%%%%%%%%%%%%%%%%%%%%%%%%%%%%%%%%%%%%%%%%%%%%%%%%%%%%%%%%%%%%%%%%%%%%%%%%%%
\section{CONCLUSIONS}
\label{sec:CONCLUSIONS}
In this paper we introduced a novel referenceless evaluation method to assess a range of audio source separation systems without the need for the original sources. We used a deep neural network to predict the sources-to-artifacts ratio (SAR) \cite{vincent:06:pmi} of singing-voice recordings extracted from music mixtures  of varying genres.
%
%In this paper we introduced a novel referenceless evaluation method to evaluate the success of separating sources using a range of audio source separation systems without the reference signals. We used deep neural networks (DNNs) to predict the sources-to-artifacts ratio (SAR) from the blind source separation evaluation toolkit (BSS-Eval) \cite{vincent:06:pmi}. 
Our experimental results show that the DNNs were capable of predicting the SAR without the reference signals, in most cases resulting in an error that was low enough (mostly $<$1.5dB) to allow discrimination between the performance of some combinations of algorithm and song, and with a high correlation (mostly $>$0.80) between the computed SAR from BSS-Eval that uses the reference signals. This work indicates that the idea of using DNNs to predict the output of objective source separation evaluation toolkits without the use of reference signals produces useful results, and can be extended to train the DNNs to predict the other metrics of the BSS-Eval or predict perceptual related quality scores. 

%%%%%%%%%%%%%%%%%%%%%%%%%%%%%%%%%%%%%%%%%%%%%%%%%%%%%%%%%%%%%%%%%%%%%%%%%%%%%%%%
\ninept
\section*{Acknowledgment}
This work is supported by grant EP/L027119/2 from the UK Engineering and Physical Sciences Research Council (EPSRC).

%{
%\vfill\pagebreak

\bibliographystyle{IEEEbib.bst}
\bibliography{refs}
%}
\end{document}